\documentclass[prl,
                twocolumn,
                preprintnumbers,
                amsmath,
                amssymb,
                showpacs,
                superscriptaddress]{revtex4-1}
\usepackage{graphicx} 
\usepackage{dcolumn}  
\usepackage{bm}       
\usepackage{amsfonts}
\usepackage[colorlinks=true,linkcolor=blue,citecolor=red,urlcolor=blue]{hyperref}
\usepackage{color}
\usepackage{amssymb}
\usepackage[dvipsnames]{xcolor}
\usepackage{soul}
\usepackage{braket}
\usepackage{amsmath}
\usepackage[normalem]{ulem}

\newcommand{\eq}[1]{Eq.\,(\ref{#1})}
\newcommand{\ful}{\mbox{C$_{60}$}}

\begin{document}

\title{Simultaneous real and momentum space electron diffraction from a fullerene molecule}

\author{Aiswarya R.}
\affiliation{%
Department of Physics, Indian Institute of Technology Patna, Bihar 801103, India}

\author{Rasheed Shaik}
\affiliation{%
School of Physical Sciences, Indian Institute of Technology Mandi, Kamand, H.P. 175075, India}

\author{Jobin Jose}
\email[]{jobinjosen@gmail.com}
\affiliation{%
Department of Physics, Indian Institute of Technology Patna, Bihar 801103, India}

\author{Hari R. Varma}
\affiliation{%
School of Physical Sciences, Indian Institute of Technology Mandi, Kamand, H.P. 175075, India}

\author{Himadri S. Chakraborty}
\email[]{himadri@nwmissouri.edu}
\affiliation{Department of Natural Sciences, D.L.\ Hubbard Center for Innovation, Northwest Missouri State University, Maryville, Missouri 64468, USA}

\begin{abstract}
{Plane-wave electrons undergo momentum transfer as they scatter off a target in overlapping spherical waves. The transferred momentum leads to target structural information to be encoded in angle and energy differential scattering. For symmetric, periodic or structured targets this can engender diffraction in the electron intensity both in real (angular) and in momentum space. With the example of elastic scattering from $\ful$ we show this simultaneous manifestation of diffraction signatures. The simulated angle-momentum diffractograms can be imaged in experiments with a two-dimensional detector and an energy-tunable electron gun. The result may inspire invention of technology to extend scopes of electron diffraction from molecules and nanostructures, open a direction of electron crystallography using the momentum-differential diffraction, and motivate an approach to control the time delay between the pump laser-pulse and the probe electron-pulse by tuning the electron impact-speed in ultrafast electron diffraction experiments.}
\end{abstract}

\maketitle


Electron diffraction (ED) studies have flourished in accessing the structural and dynamical information of molecules~\cite{saha2022,amini2020}, nanostructures~\cite{ponce2021}, surfaces~\cite{Kienzle2012}, and crystals~\cite{gemmi2019}. In standard ED measurements, the target is irradiated by an electron beam of a fixed energy and the scattered intensity is measured by a two-dimensional (2D) detector. This detector can be either a luminescent, namely phosphor, screen coupled to a charge-coupled device or a direct hit detector that collects diffracted signal within a large solid angle. For crystals, where the sample orientation can also be tuned, measurements are typically performed in a transmission electron microscope which provides access to targets otherwise inaccessible by X-ray diffraction (X-rays are blind to the Coulomb force). Ultrafast ED (UED)~\cite{centurian2022} is a popular technique which temporally resolves the evolutionary dynamics of an excited sample. In this, the sample is initially excited by a laser pulse (pump) and then interrogated by a time-delayed electron pulse (probe). The ED signal is registered by varying pump-probe delay in a controllable manner. The fundamental notion that a diffraction pattern encodes information about the Fourier transform of the real-space structure of the target that produced it, and {\em vice versa}, is at the heart of these studies.

Diffraction is the bending of an incident plane wave into effective spherical waves, at detectable distances larger than the target size, by interacting with the structural details of the target. The resulting interference produces patterns in the signal. Fundamentally, therefore, the pattern is a function of the magnitude ($q$) of transferred momentum vector, the difference between the final and the initial momentum of the projectile. Let us consider the elastic scattering for simplicity in which the electron energy, and therefore the magnitude of the electron momentum ($\vec{k}$), does not alter during the scattering. In this case, $q$ for a scattering angle $\theta$ can be written as,
\begin{eqnarray}\label{tran-mom}
q &=& 2k\sin(\theta/2).
\end{eqnarray}
\eq{tran-mom} implies that one way to vary $q$ is to keep the electron energy fixed but capture the signal at various angles. This is routinely done in current ED measurements by using a 2D detector. This provides the {\em fringe} pattern in the real ($\theta$) space. An alternative way, see \eq{tran-mom}, to scan the $q$-space for a fixed detection angle is to vary the impact energy to generate diffraction as a function of $k$. An advantage of the latter is that with increasing energy and therefore decreasing de Broglie wavelength the electrons can resolve finer details of the target structure – from the larger-scale electron density to the smaller-scale ion-core and inter-nuclear structures. For UED, the tuning of impact energy may directly control the pump-probe delay time, in particular to resolve dynamics from picoseconds~\cite{centurian2010,yang2015} to nanoseconds time-scale, the latter being prevalent in the field of atmospheric chemistry, biology and combustion science~\cite{long2023}. This can be achieved by locking on a fixed launch-time difference of the pulses, while controlling their arrival-time delay as a function of the electron speed. In general, by combining both these pathways (see schematic Fig.\,1), namely by using an energy-tunable source and a 2D detector, it will be possible to generate a 2D diffraction map, the diffractogram, in combined real-and-momentum space (Fig.\,5).
\begin{figure*}[t]
\includegraphics[width=0.9\textwidth]{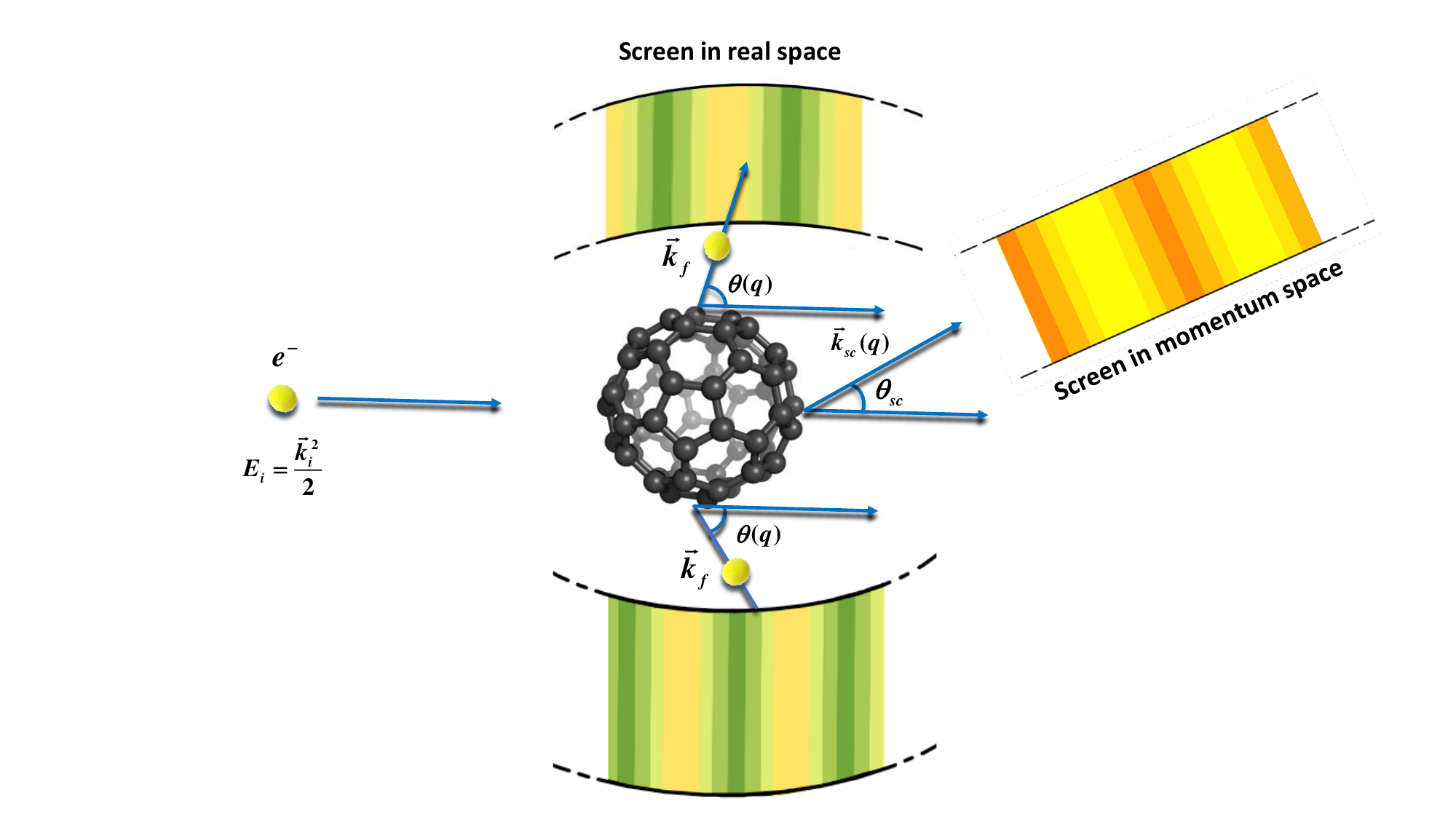} 
\caption{(Color online) A cartoon to delineate the elastic electron impact diffraction with $\ful$. It schematically shows diffraction fringes in scattered intensity as a function of the scattering angle $\theta$ for a fixed incident energy and the fringes at a given scattering direction $\theta_{\scriptsize sc}$ as a function of the magnitude of electron momentum $k = |\vec{k}_i| = |\vec{k}_f|$ as the incident energy varies.}
\label{fig1}
\end{figure*}

In this study, we calculate the elastic electron scattering differential cross section (DCS) of a $\ful$ molecule target and produce such diffractograms. The study serves as a proof-of-principle to provide impetus in accessing the entire ED landscape by including the momentum space information. This may motivate extensions of the ED method's general spectroscopic scope. Since the discovery of $\ful$~\cite{kroto85}, this highly symmetric molecule has been drawing extensive research attention, both for fundamental reasons as well as for its broad applied relevance~\cite{full-appl2021}. The high symmetry also makes it an efficient diffracting target. Experiments on the electron scattering from $\ful$ have included low-energy~\cite{tanaka2021}, intermediate-energy~\cite{hargreaves2010}, and high-energy~\cite{garchekov1998} measurements. It is generally accepted that the e-$\ful$ collision DCS exhibits diffraction effects by revealing minima and maxima as a function of $\theta$ and this pattern differs for various scattering energies~\cite{tanaka2021,hargreaves2010,garchekov1998}. The diffraction behavior of elastic/inelastic high energy DCS was found to connect to the geometry of $\ful$ electronic distribution~\cite{garchekov1998,garchekov1997}. However, little assertion of the $k$-space diffraction effect is made to the best of our knowledge. Further, even though some qualitative attributes are drawn, the quantitative relation of the diffraction in DCS to the target geometry is not demonstrated, which we include in the current study.

There is a distinction in how the diffraction is probed in e-$\ful$ scattering process versus in photoionization (PI) and positronium (Ps) formation with $\ful$, in both of which the electron exists only in the final channel (half-scattering for PI). PI and Ps formation are intrinsically inelastic, leading to the variation of the momentum of the observed particle, photoelectron or Ps, as the impact energy varies. This is true for the detection of the particle at a certain angle, integrated over a small angular range in the forward direction, or integrated over all angles. Consequently, the diffraction phenomenon is accessible in the momentum space in these processes. Indeed, the $k$-space diffraction in the PI from highest occupied molecular orbitals of $\ful$ were measured and calculated~\cite{ruedel2002,mccune2008}. Also, diffraction in the angle-integrated~\cite{hervieux2017} and forward direction~\cite{hervieux2019} Ps formation in positron-$\ful$ collisions was predicted where the feature was captured in the Ps momentum. But for PI of predominantly spherical systems, like the valence electron cloud in $\ful$, the diffraction in $\theta$ is forbidden~\cite{AngDistPI,Cooper1993}. This notion is similar to the classical Fraunhoffer type diffraction through a spherical aperture~\cite{guasti93}. For the Ps formation, even though $q$ of the incoming positron is involved, it gets non-trivially entangled with $k$ of the captured electron~\cite{hervieux2017}. On the other hand, as pointed out above, the fundamental domain to map out diffraction in elastic scattering is by varying $q$ of the electron. The information, thus, automatically translates into the variation either of $\theta$ (a direct quantum mechanical analogue of classical Young’s diffraction) or of $k$, enabling a {\em complete} description of the process. This is also a novelty of the current study. 

The details of the computational methods are included as Supplementary Material (SM)~\cite{SM}. The isolated $\ful$ molecule is modeled by a jellium-based local-density approximation scheme~\cite{madjet2008} within density functional theory (DFT). A spherical frame is employed that includes 240 delocalized $\ful$ electrons. Variants of this DFT model were successfully employed in explaining measurements of (i) photoelectron~\cite{ruedel2002} and photoion~\cite{scully2005} intensity, respectively, at non-plasmonic and plasmonic energies, and (ii) a recent plasmonic PI time delay spectra~\cite{biswas2022}. A simple electron-less (static) annular square well (ASW) potential to model the $\ful$ shell with parametric adjustment to match the electron affinity of $\ful$~\cite{winstead2006}, 2.65 eV known from the experiment \cite{tosatti1994anomalous}, is used in parallel to identify robust features. These potentials mediate the e-$\ful$ interaction, that further includes an appropriate polarization potential~\cite{SM} for a fuller description. The e-$\ful$ elastic scattering parameters are then obtained using the non-relativistic partial wave technique. The radial Schr\"{o}dinger equation is numerically solved to access the scattering state of energy $E$ and, subsequently, the scattering phase shift $\delta_{\ell}$ with corresponding angular momentum quantum number $\ell$. The total scattering amplitude for a scattering angle $\theta$ measured from the incident (forward) direction is then evaluated as,
\begin{eqnarray}\label{amp}
f(k,\theta) &=& \frac{1}{2ik}\sum_{\ell=0}^{\infty}(2\ell+1)P_\ell(\cos \theta)\left[\exp(i2\delta_\ell) - 1\right],
\end{eqnarray}
where the magnitude of the projectile momentum, in atomic units (a.u.), $k=\sqrt{2E}$ and $P_\ell$ is the Legendre polynomial. The DCS is then evaluated as,
\begin{eqnarray}\label{dcs}
\frac{d\sigma}{d\Omega} &=& |f(k,\theta)|^2.
\end{eqnarray}
\begin{figure}[h!]
\includegraphics[width=9.0 cm]{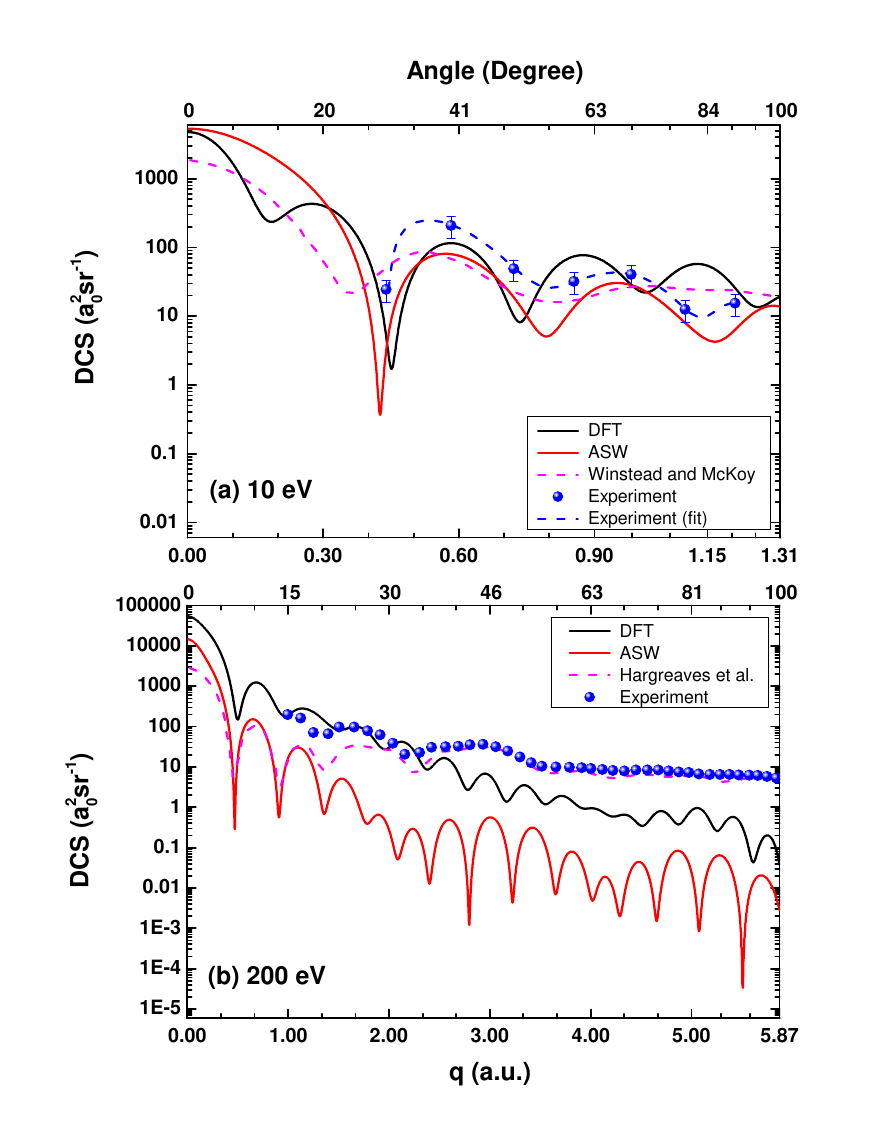} 
\caption{(Color online) Angle-differential DCS for e-$\ful$ elastic scattering calculated in DFT and ASW models and compared with measurements~\cite{tanaka2021,hargreaves2010} and SMC calculations~\cite{winstead2006}. Both $q$ and $\theta$ scales are shown. Two impact energies considered are 10 eV (a) and 200 eV (b).}
\label{fig2}
\end{figure}

In Fig.\,2 we compare our results of the angle-differential DCS with measurements and other calculations for two values of the impact energy. Note that the $q$-scale is also shown. For the lower 10-eV impact energy, panel (a), both of our DFT and ASW results agree reasonably well with overall features of the measurements~\cite{tanaka2021}. The agreement achieved by a successful and accurate Schwinger multi-channel (SMC) method~\cite{winstead2006} is of the same level of our calculations, as also shown. We note that SMC includes an atomistic model of $\ful$ and therefore has richer structural information than a jellium $\ful$. In spite of this, the success of our DFT results versus measurements and SMC for a low-energy impact likely indicates that longer de\,Broglie wavelength of the slower electron can not resolve atomistic structures. 

For the faster 200-eV impact, panel (b), while the ASW result fails to agree with the experiment~\cite{hargreaves2010}, DFT fairs well versus measurements at lower $q$ values around the forward scattering. At larger scattering angles, however, DFT not only shows more structures than what are present in the measured data, but also weakens in the overall intensity. On the other hand, the SMC calculation~\cite{hargreaves2010} compares well with the experiment at wider angles, while not so well toward the forward scattering. In general, these structures in DCS are qualitatively known to originate from the ED effect of $\ful$. However, there are other effects, particularly for higher impact energies and backward scattering, namely the direct collisions with individual carbon atomic centers, which the jellium DFT and ASW models do not include. On the other hand, this atomic contribution at larger scattering angles may wash out diffraction effects which is likely the case for measurements and SMC. Also, the wavelength of 200-eV electrons is not short enough to diffract from the C-C bond ($\sim$ 0.15 nm). We remark that since the soft-edged DFT potential can scatter electrons from its long-range ``wings" on either side of $\ful$ shell (Fig.1 in SM), while the much harder-edged ASW potential cannot, the overall intensity of DCS in the former is much stronger.
\begin{figure}
\includegraphics[width=9.0 cm, keepaspectratio]{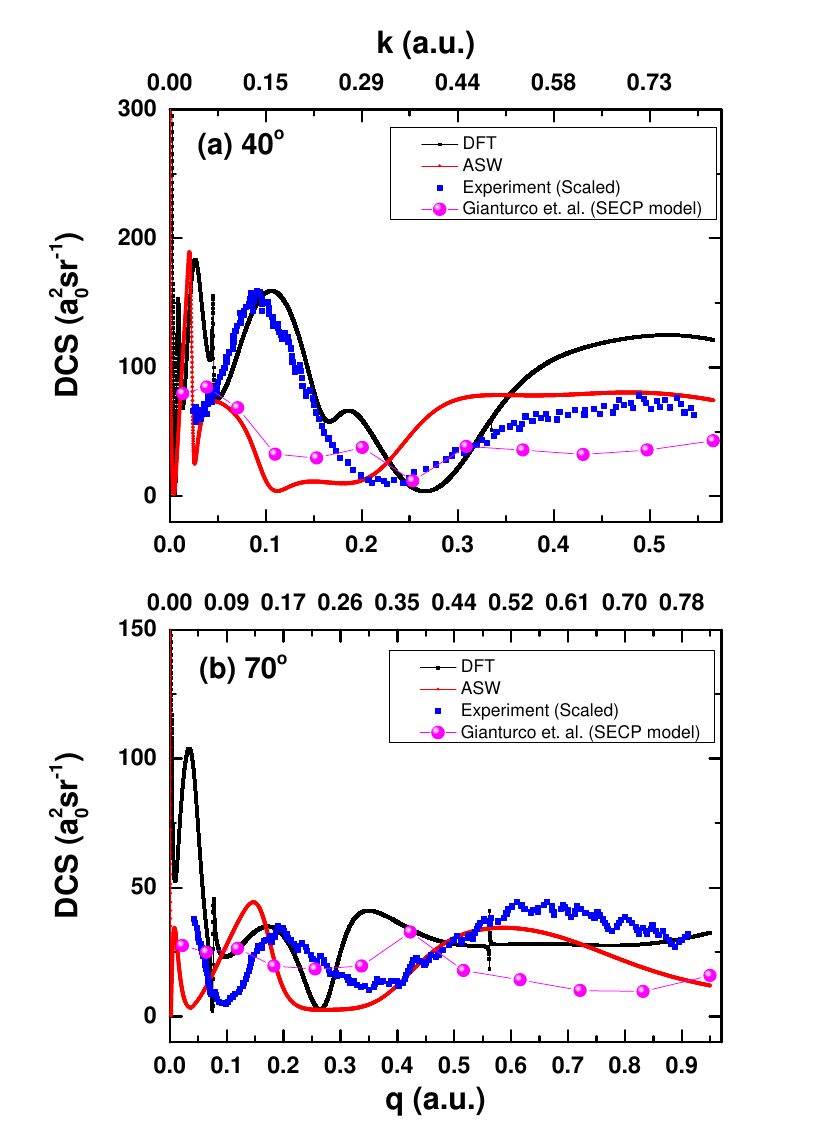} 
\caption{(Color online) Momentum-differential DCS for e-$\ful$ elastic scattering calculated in DFT and ASW models, and compared with measurements~\cite{tanaka1994} and SECP calculations~\cite{gianturco1999}. Both $q$ and $k$ scales are shown. Two scattering angles  considered are 40$^\circ$ (a) and 70$^\circ$ (b).}
\label{fig3}
\end{figure}

In Fig.\,3 we display our momentum-differential DCS, and compare with measurements and another theory, for two choices of the scattering angle. For the 40$^\circ$ scattering on panel (a), our DFT results show a very good agreement with the measurements~\cite{tanaka1994}. This is particularly remarkable given that the static exchange correlation polarization (SECP) calculations~\cite{gianturco1999} (also presented) significantly disagree with the experiment. Further, ASW also showing a poor agreement suggests the superiority of DFT at this angle. Panel (b) exhibits results for 70$^\circ$. Both DFT and ASW qualitatively capture structures in the measured profile~\cite{tanaka1994}, while SECP still somewhat struggles to match. In any case, the structures in $k$-differential DCS in Fig.\,3 must also be due to the diffraction mechanism, since fundamentally ED is a function of $q$; the $q$ variation is shown in the figure. Therefore, Figs.\,2 and 3 together embody a notion of complete manifestation of ED that we will scrutinize further in the following.


\begin{figure}
\includegraphics[width=9.0 cm, keepaspectratio]{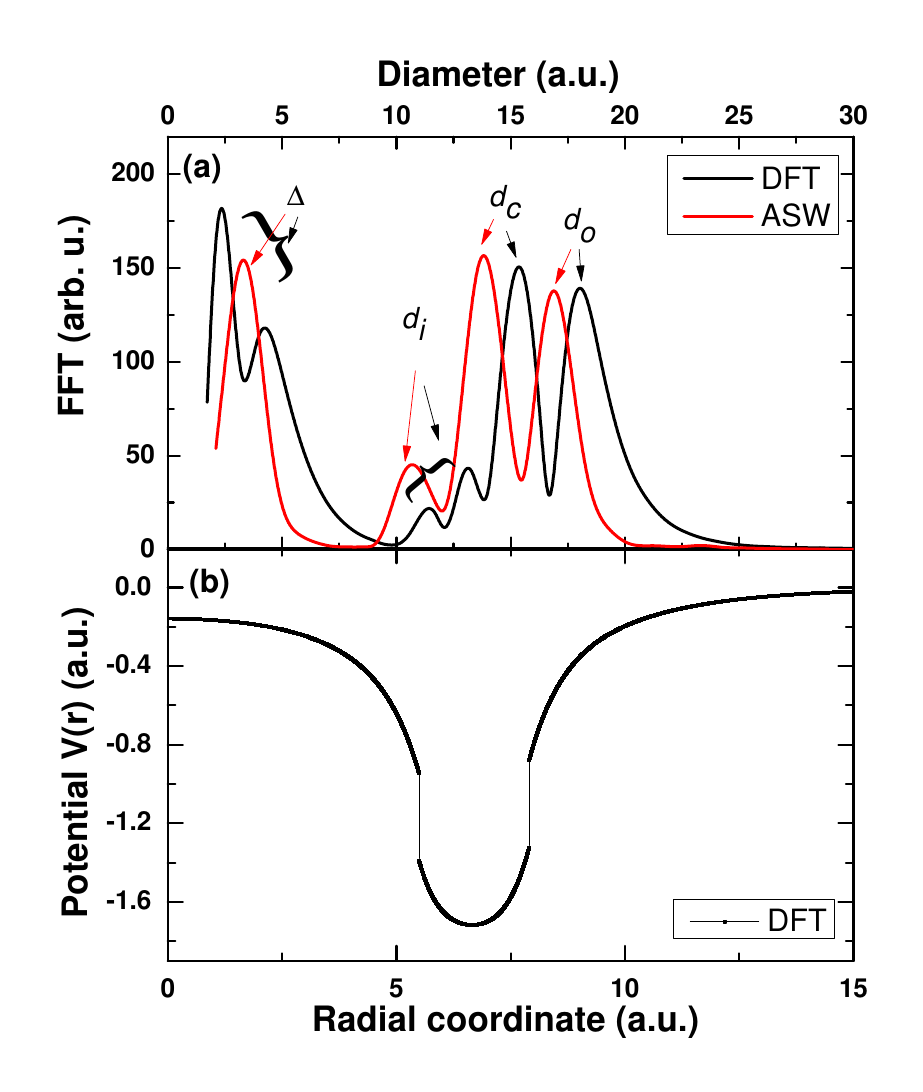} 
\caption{(Color online) Fast Fourier transforms (FFT) of e-$\ful$ elastic scattering DCS, computed in ASW and DFT schemes (a). The peaks are attributed to $\ful$ geometric features (see text). The DFT radial interaction potential is plotted (b) to facilitate the geometric comparison.}
\label{fig4}
\end{figure}
The Fourier transform of the diffraction signal in scattered electrons maps out the diffractor's structural information that can be resolved by the wavelength of the incoming electron. In order to assess this, we first benchmark the expectation by using the ASW potential in the first Born approximation of elastic scattering to analytically evaluate DCS as a function of $q$~\cite{SM}. The result indicates that the diffraction fringes in DCS, or equivalently the intensity ``waves" as a function of $q$, carry the ``frequencies" corresponding to $\ful$ shell structure. These frequencies are: the $\ful$ shell's inner ($d_{\scriptsize i}$) and outer ($d_{\scriptsize o}$) diameters, the average molecular diameter $d_{\scriptsize c}=(d_{\scriptsize i}+d_{\scriptsize o})/2$, and the shell width $\Delta =(d_{\scriptsize i}-d_{\scriptsize o})/2$. The core mechanism includes diffraction from the inner and outer edge of the shell in the scattering amplitude \eq{amp} and their interference by the coherence process (squared modulus) in DCS \eq{dcs}. These frequencies must be found in the reciprocal spectra of the DCS.

Therefore, we analyze numerically computed DCS as a function of $q$ by carrying out their Fourier transform (FT). This is done by a fast FT (FFT) algorithm after neutralizing the steady background decay of the DCS and appropriately windowing the data before FFT feed~\cite{mccune2008}. Obtained FFT signals are presented in Fig.\,4(a) that indeed reveal peaks corresponding to $\ful$ geometry. The ASW result better captures the positions of the edges when compared with panel (b). However, the DFT result shows offsets and some splits in the frequency peaks. This is due to the fact that the shell edges are somewhat diffused in the DFT interaction potential [Fig.\,4(b)]. In effect, the diffraction in the DFT approach maps out the delocalized electron density-profile (around the molecular shell) instead, which is realistic.

In order to image the complete landscape of the diffraction phenomenon, we generate iso-surface 2D polar maps, the diffractrograms, of the DCS in combined real ($\theta$) and momentum space in Fig.\,5. The $\theta$ range is chosen between $-120^{\circ}$ and $120^{\circ}$, because the signal around the back-scattering is found too weak. The primary fringe pattern exhibits series of parabolas. One finds an overall impression of weaker groupings of fringes, more obvious in ASW (lower panel), which denotes superimposed wider (beating) oscillations connected to the width $\Delta$, which is the smallest size feature in FT [Fig.\,4(a)]. Note further that the DFT image (upper panel) reveals additional parabolas towards back-scattering. These secondary fringe pattern originate from the even-function symmetry, $\cos(\theta) = \cos(\pi-\theta)$, applied in \eq{amp}. Both the patterns overlap for a range of $\theta$, as seen. However, the intensity of the patterns depends on the sum of the Legendre polynomials in \eq{amp} and their interference in \eq{dcs}. It is well known that Legendre polynomials of increasing order $\ell$ oscillate progressively in phase toward the forward direction producing a constructive interference, resulting significantly stronger primary-fringe intensity. But the polynomials' out-of-phase oscillations inducing destructive interference to the backward direction weaken the secondary-fringe intensity. However, since with increasing $k$ more partial waves are included, this effect enhances causing an effective extinction of secondary fringes for higher $k$ as noticed. Obviously, the effect is physical in the more realistic DFT approach and may be detectable in experiments. We reiterate, similar diffractograms can be generated in experiments using energy-tunable electron source and a 2D detector. As an attractive prospect for contemporary ultrafast studies, similar angle-momentum differential time-delay diffractograms can be accessed; temporal fringe patterns were predicted for photoelectrons~\cite{magrakvelidze2015}
\begin{figure}
\includegraphics[width=10.0 cm]{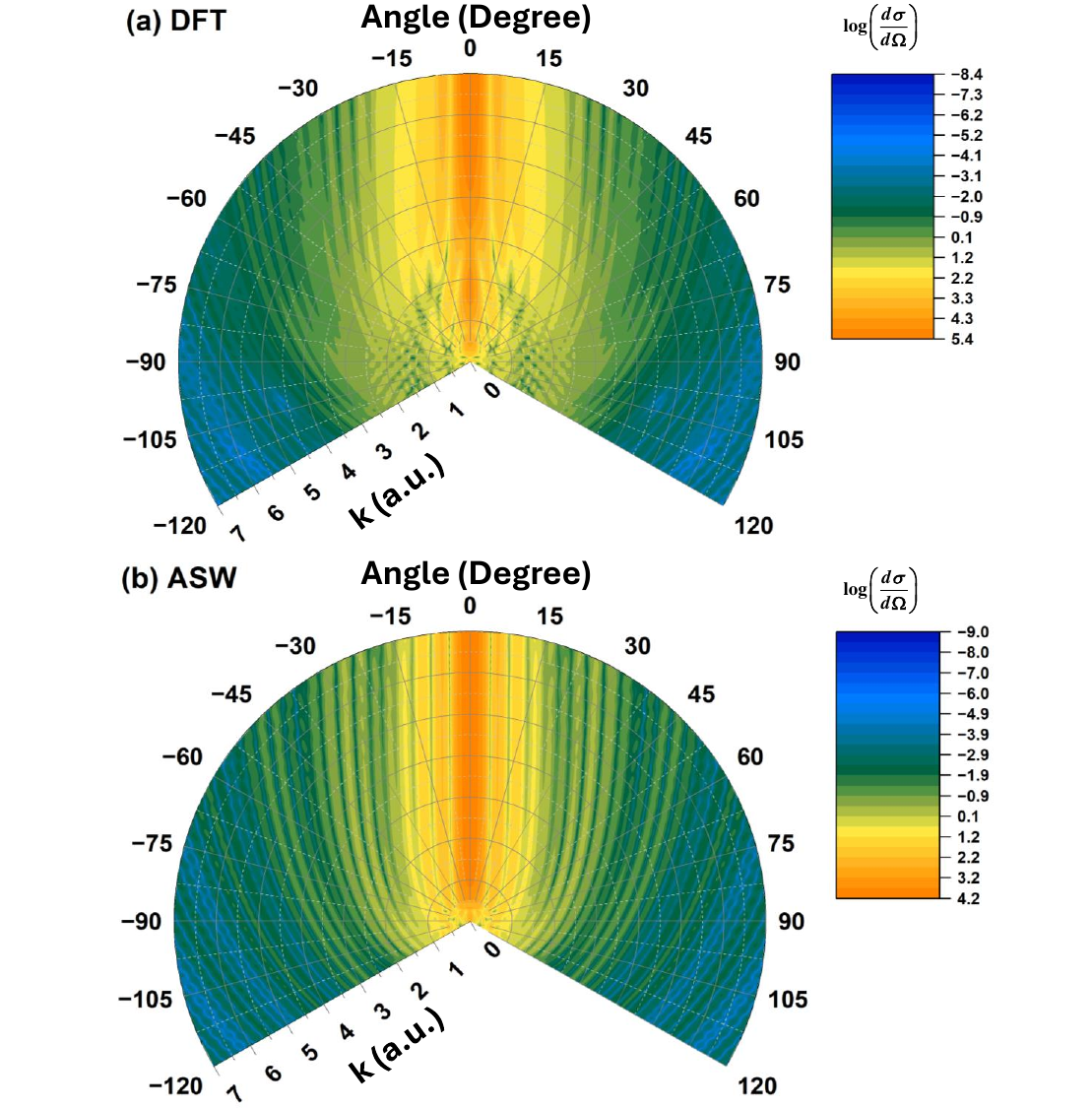} 
\caption{(Color online) Angle-momentum combined 2D iso-surface polar-images of e-$\ful$ elastic scattering DCS with the DFT potential (top) and ASW potential (bottom) showing rich diffraction features.}
\label{fig5}
\end{figure}

In conclusion, considering the electron elastic collision with $\ful$ we demonstrate that combined real and momentum space 2D DCS maps out the complete diffraction topography of scattering. An advantage of accessing the momentum space, or equivalently tuning the electron's wavelength, is to find a knob for increasing resolution of the structure. We used a fairly successful DFT model of $\ful$, in parallel with a simpler model to hone insights. The computed results show good agreements with available measurements, often achieving better success than existing theories. The Fourier transform of the DCS is invoked for the first time to reveal structural information which is the capstone goal of ED studies of molecular or crystal targets. The diffractogram simulated on the DFT track is found rich in spectroscopic information. Such images can be generated in {\em complete} experiments with gas phase $\ful$. The study may also motivate experiments for fullerenes in solutions or as thin films. Even though the collision may also generate electrons by various inelastic channels including the $\ful$ plasmon channel, they can be selected out by appropriate energy analysis.

Moreover, there is a broader implication of the study. This may encourage the extension of the current technology in order to explore the momentum space electron scattering from a molecule in gas phase and electron transmission through a crystal. This will tap into structural information inaccessible in constant irradiation-energy mode. The study can also inspire the electron-speed controlled time delay of the probe electron pulse from the pump laser pulse in UED approaches to molecular dynamics. It can be estimated that within the irradiation energy range from 100 keV to 100 eV one may tune the electron pulse's arrival delay from picoseconds to nanoseconds ranges with 10 to 100 eV energy increments. There are many important chemical and bio-molecular processes occur within these time scales. In addition, the evolution from nonequilibrium nuclear dynamics from the sub-picosecond timescale~\cite{srinivasan2003,dudek2001} to slower nanoseconds ground-state kinetics through microcanonical energy distribution pathways is still not clearly known. To this end therefore, the current study may offer a handle to the control of the time resolution while simultaneously upgrading the structure resolution to impact ultrafast molecular science.  

\begin{acknowledgments} 
We thank Dr. Martin Centurian for helpful discussions. The research is supported by CRG/2022/000191, India (JJ), DST-SERB-CRG Project No. CRG/2022/002309, India (HRV), and US National Science Foundation Grant No. PHY-2110318 (HSC.).
\end{acknowledgments}


\end{document}